\documentclass[a4paper,10pt]{article}
\usepackage{graphicx}
\usepackage{mathtools} 
\usepackage[left=2.2cm, right=2.2cm, top=3cm]{geometry}
\pdfoutput=1

\title{Optimization of detector modules for measuring gamma-ray polarization in Positron Emission Tomography\footnote{Preprint submitted to Nuclear Instruments and Methods in Physics Research Section A} } % Article title

\author{Siddharth Parashari\textsuperscript{1}\footnote{Corresponding author: siddharth@phy.hr}, Mihael Makek\textsuperscript{1}, Tomislav Bokuli\'{c}\textsuperscript{1}, Damir Bosnar\textsuperscript{1}, \\Ana Marija Ko\v{z}uljevi\'{c}\textsuperscript{1}, Zdenka Kuncic\textsuperscript{2}, Petar \v{Z}ugec\textsuperscript{1}} % Authors
\date{}

\begin{document}
\let\WriteBookmarks\relax
\def\floatpagepagefraction{1}
\def\textpagefraction{.001}

\maketitle % Print the title and abstract box

\begin{flushleft}
\textsuperscript{1}\textit{Department of Physics, Faculty of Science, University of Zagreb, Bijeni\v{c}ka c. 32, 10000 Zagreb, Croatia} \\ % Author affiliation 
\textsuperscript{2}\textit{School of Physics, University of Sydney, New South Wales, Sydney, 2006, Australia}  \\ % Author affiliation
\end{flushleft}

%----------------------------------------------------------------------------------------
%	ARTICLE CONTENTS
%----------------------------------------------------------------------------------------

\begin{abstract}
 Detection of $\gamma$-ray polarization in Positron Emission Tomography (PET) is as yet an unexploited feature
that could be used as an additional handle to improve signal-to-background ratio in this imaging modality.
The $\gamma$ polarization is related to the azimuthal angle in the Compton scattering process, so the initial
correlation of polarizations of the annihilation quanta translates to the correlation of the azimuthal angles in events where both annihilation photons undergo Compton scattering. This results in a  modulated distribution of the azimuthal angle difference for true events, while this modulation is lacking for the background events. We present a comprehensive experimental study of five detector configurations based on scintillator matrices and silicon photomultipliers, suitable for measuring the azimuthal modulation. The modules consist of either GaGG:Ce or LYSO:Ce pixels with sizes varying from 1.9x1.9x20 $\mathrm{mm^3}$ to 3x3x20 $\mathrm{mm^3}$. The distinctive feature of the modules is that they can reconstruct the Compton scattering by detecting the recoil electron
and the scattered gamma in a single detector layer, which simplifies extension to larger systems. The amplitude modulation of the azimuthal angles' difference is clearly observable in all configurations ranging from $0.26\pm0.01$ to $0.34\pm 0.02$ depending on the event  selection criteria. The results suggest that finer detector segmentation plays a leading role in achieving higher modulation factors.
\end{abstract}


\section{Introduction}
Detection of gamma-ray polarization in Positron Emission Tomography (PET) gained a rising interest in
recent years, primarily driven by simulation studies suggesting it can improve signal-to-background ratio
in this imaging modality \cite{1,2,3}. The benefit of the polarization measurement comes from the fact that the
511 keV quanta created after positronium annihilation are entangled and they have orthogonal
polarizations. The polarization is strongly related to the azimuthal angle in the Compton scattering process,
therefore the initial correlation of polarizations translates to the correlation of azimuthal angles. Clearly,
background events in PET lack such correlation, so they can be discriminated from true coincidences.
Experimentally, the detection of azimuthal correlations poses a challenge, mainly because the full
reconstruction of Compton events requires position sensitive devices which can detect the recoil electron and the scattered gamma. If separate detection layers are be used to detect them, it would significantly increase the size, complexity and cost of a clinical PET system. A proof-of-concept study demonstrated that detection of the azimuthal correlations is feasible by using single matrices of scintillator detectors read out on one side by silicon photo-multipliers (SiPM) \cite{4}. In the present work, we have as well studied the single-layer detectors to achieve the optimum azimuthal modulation using five detector configurations of either GaGG:Ce or LYSO:Ce pixels with varying pixel sizes from $1.9\times1.9\times20 \;\mathrm{mm^3}$ to $3.0\times3.0\times20 \;\mathrm{mm^3}$. The goal of the study is to compare the studied configurations and to determine the prospective candidates for implementation in a full-scale PET system.

\section{Gamma polarization measurement via Compton scattering}
Two photons of 511 keV energy emitted back-to-back from $\mathrm{e^+e^-}$ annihilation events are orthogonally polarized.
In case both of them undergo Compton scattering with scattering angles $\theta_{1,2}$ and azimuthal angles $\phi_{1,2}$, respectively, the differential cross-section is given by \cite{5},
\begin{equation}
\mathrm{ \frac{d^2\sigma}{d\Omega_1\!d\Omega_2}\!\!=\!\!\frac{r_0^4}{16}F(\theta_1\!)F(\theta_2\!)\left[1\!-\!\frac{G(\theta_1\!)G(\theta_2\!)}{F(\theta_1\!)F(\theta_2\!)}cos[2(\phi_1\!\!-\!\phi_2\!)]\right] }
\end{equation}
with $\mathrm{F(\theta_i)=\frac{2+(1-cos\theta_i)^3}{(2-cos\theta_i)^3}}$ and $\mathrm{G(\theta_i)=\frac{sin^2\theta_i}{(2-cos\theta_i)^2}}$, where i=1,2.
For the fixed values of $\theta_{1,2}$, the cross-section has maxima for $|\phi_1-\phi_2|=90^\circ$. This reflects the fact that the polarization correlation is dominantly preserved when both gammas undergo Compton scattering, since they were initially orthogonally polarized. 

The sensitivity of the measurement to the initial relative polarization of the gamma
photons can be quantified by the polarimetric modulation factor ($\mu$),
\begin{equation}
   \mathrm{ \mu=\frac{P(\phi_1-\phi_2=90^\circ)-P(\phi_1-\phi_2=0^\circ)}{P(\phi_1-\phi_2=90^\circ)+P(\phi_1-\phi_2=0^\circ)} }
\end{equation}
where, $\mathrm{P(\phi_1\!-\!\phi_2=90^\circ)}$ and $\mathrm{P(\phi_1\!-\!\phi_2=0^\circ)}$ are the probabilities of observing the two gammas scattering with orthogonal or parallel azimuthal angles, respectively. The expected modulation is the strongest for $\theta_1\!=\!\theta_2\!=\!82^\circ$ and its maximum value is $\mu_{max}\!=\!0.48$ \cite{5}. Therefore, the events with the scattering angles close to $\theta_{1,2}\!=\!82^\circ$ are the most relevant for discriminating the signal and the background and the higher the modulation amplitude, the better discrimination will be possible.
For realistic detector geometries, the values of $\mu$ are lower than  $\mu_{max}$ since they are averaged over the finite detector acceptances, which are not only governed by the selected kinematic range, but also by the energy resolution and the angular resolution of the detectors. In this study we compare the polarimetric performance of the detectors with different energy resolutions (see Table \ref{tbl1}) and with inherently different azimuthal resolutions, driven by the detector segmentation.

\section{Methodology}
\begin{figure}
	\centering
		\includegraphics[width=9.5cm]{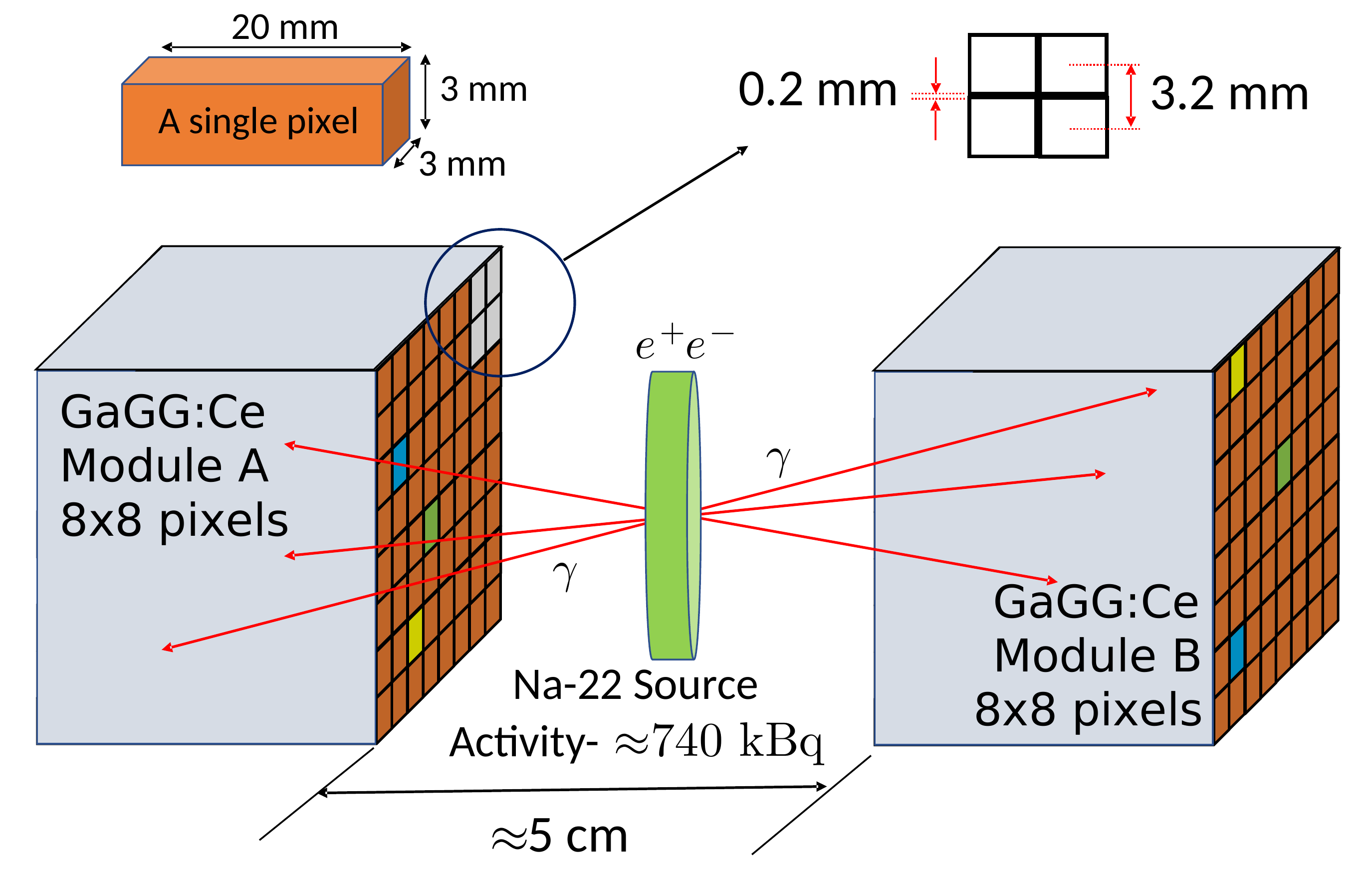}
	\caption{Schematic diagram of experimental setup example for GaGG-3.0 configuration}
	\label{FIG:1}
\end{figure}

A system of two identical modules, approximately 5 cm apart, with a $\mathrm{^{22}Na}$-source (diam. 1 mm, activity $\!\approx$ 740 kBq) placed in the middle, is used for the measurements (Figure \ref{FIG:1}). Five different module configurations have been set up and tested, as listed in the Table \ref{tbl1}. In each configuration, a module comprises a 8x8 matrix of scintillator pixels read-out by a SiPM array with one SiPM pixel matching one crystal in the matrix. The SiPMs are optically coupled to the crystal matrices using optical cement (EJ-500, Eljen Technology, USA), except in the GaGG-3.0 configuraton where one module is coupled using silicon optical pad \cite{6}. All the measurements have been carried out at an over voltage, $\mathrm{V_{OV}=4.0\; V}$ at temperatures $18\pm1\;^\circ C$. The spectrum obtained from each pixel is individually calibrated and corrected for non-linearity \cite{6}. The obtained mean energy resolutions at 511 keV from different detectors are listed in Table \ref{tbl1}, where the uncertainty reflects the energy resolution spread of pixels within the modules. The coincidence data are acquired using the TOFPET2 system with the default configuration \cite{7}.

\begin{table}
\caption{Detector module configurations and their respective properties.}
\centering
\label{tbl1}
\begin{tabular}{clllcc} \hline

Setup & Modules & Crystal & Pixel & Pitch & Resolution\\
     & & Material & Dim. ($\mathrm{mm^3}$) & ($\mathrm{mm}$) & (keV)\\ \hline

1 & GaGG-1.9 & GaGG:Ce & 1.9x1.9x20 & 2.2 & $8.1\pm0.5$\\
2 & GaGG-2.9 & GaGG:Ce & 2.9x2.9x20 & 3.2 & $8.5\pm0.5$\\
3 & GaGG-3.0 & GaGG:Ce & 3.0x3.0x20 & 3.2 & $10.2\pm0.8$\\
4 & LYSO-2.0 & LYSO:Ce & 2.0x2.0x20 & 2.2 & $13.7\pm0.9$\\
5 & LYSO-1.9 & LYSO:Ce & 1.9x1.9x20 & 2.2 & $14.8\pm1.2$\\ \hline
\end{tabular}
\end{table}

\begin{figure}
	\centering
		\includegraphics[width=8.5cm]{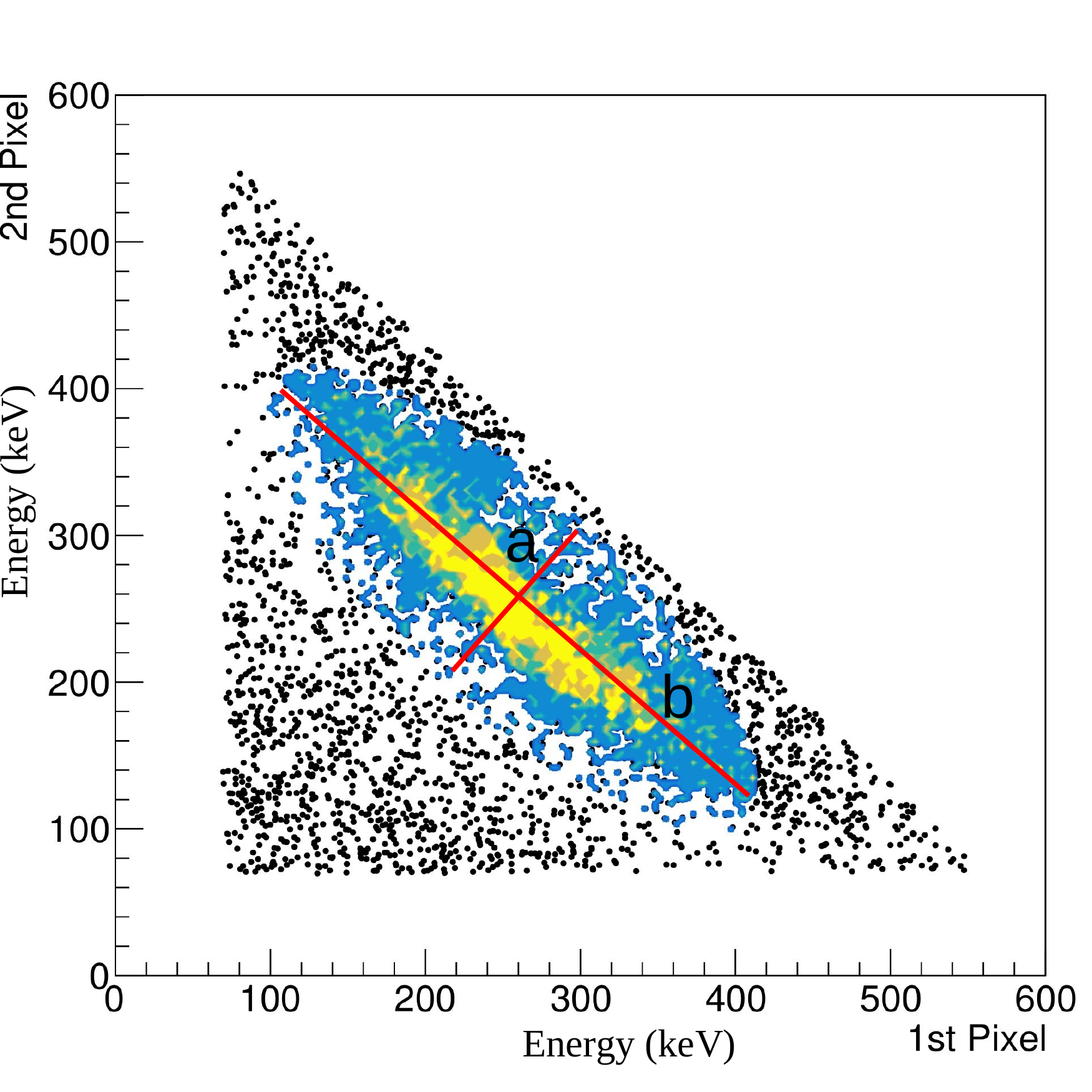}
	\caption{Selected Compton events (color region). }
	\label{FIG:2}
\end{figure}
The Compton events that occur in a detector module are selected requiring that exactly two pixel fire per module and that the sum of their energies is within $\pm3\sigma$ from the 511 keV peak maximum ($\mathrm{E_{\gamma}}$). A lower bound of 100 keV is applied to all pixel energies to avoid possible noise contributions. Additional criteria corresponding to Compton scattering kinematics are used to select a clean sample of Compton events in each module, as depicted in Figure \ref{FIG:2}, with the following condition,
\begin{equation*} 
   \mathrm{ \left(\frac{E_{px_1}+E_{px_2}-E_{\gamma}}{\mathit a} \right)^2+\left(\frac{E_{px_1}-E_{px_2}}{\mathit b} \right)^2<1 } \nonumber
\end{equation*}
where, $\mathrm{E_{px_1}}$, $\mathrm{E_{px_2}}$ are the energies of the two fired pixels and {\it a} and {\it b} are the minor and major axis for the cut. The selected value of {\it a} corresponds to $3\sigma$ of the 511 peak width in the module. The value of {\it b} corresponds to the selected Compton energy range after applying the lower energy bound.
The Compton scattering angle ($\theta$) and the azimuthal angle ($\phi$) in each module are deduced as: 
\begin{equation}
\mathrm{
\theta = acos\left(\frac{m_ec^2}{E_{px_1}\!\!+\!E_{px_2}}-\frac{m_ec^2}{E_{{px}_2}}\!-\!1\right)\!; \phi = atan\left(\frac{\Delta y}{\Delta x}\right)
}
\end{equation}
where we assume that the first interaction happens in the pixel with the lower energy corresponding to the recoil electron ($\mathrm{E_{px_1}=E_{e'}}$, $\mathrm{E_{px_2}=E_{\gamma'}}$), since the cross-section and the detector configuration favor forward scattering \cite{8}. The $\phi$ angle is reconstructed from $\mathrm{\Delta x}$ and $\mathrm{\Delta y}$, the distances of the fired pixel centres in the x-y plane (perpendicular to the longer crystal axis). The uncertainty in determination of scattering angle, $\theta$, depends on the energy resolution of the pixels and following the principles of error propagation we obtain the resolution of the  scattering angle ($\mathrm{\Delta\theta}$) within $15.3^\circ-29.6^\circ$ (FWHM) for the setups 1-5, respectively. These uncertainties remain almost constant over the range of the reconstructed scattering angle $50^\circ\!<\!\theta<90^\circ$ \cite{9}. The azimuthal uncertainty for each Compton event is calculated as $\sigma_\phi=a/{(d\sqrt{6})}$, where $a$ is the cross-sectional side of the pixel and $d$ is the distance $d=\sqrt{\Delta x^2 + \Delta y ^2}$ of the fired pixel centres. The mean azimuthal resolution, $\left<\Delta\phi\right>$, is obtained as the average resolution ($2.35\, \sigma_\phi$) of all selected events \cite{9}.

To quantify the azimuthal modulation, we select events in which both quanta undergo Compton scattering and reconstruct the distribution of the azimuthal angle differences, $\mathrm{N(\phi_1-\phi_2)}$. This distribution is then corrected for acceptance, as described in \cite{4}. The modulation factor, $\mu$, is determined by fitting the acceptance-corrected distribution, $\mathrm{N_{cor}(\phi_1-\phi_2)}$, with:
\begin{equation}
    \mathrm{N_{cor}(\phi_1-\phi_2)=M[1-\mu \; cos(2(\phi_1-\phi_2)]}
\end{equation}

\begin{table}
\caption{Modulation factors $\mu$ and mean azimuthal resolutions $\left<\Delta\phi\right>$ (FWHM) at three different ranges of $\theta_{1,2}$. Condition of inter-pixel distance $\mathrm{d_{1,2}\!>\!4.5}$ mm is applied. The uncertainties shown in $\mu$ are statistical. The systematic uncertainties are discussed in the text.}
\label{tbl2}
\centering
\begin{tabular}{cccccccc}
\hline
& & \multicolumn{2}{c}{$80^\circ<\theta_{1,2}<84^\circ$} & \multicolumn{2}{c}{$77^\circ<\theta_{1,2}<87^\circ$} & \multicolumn{2}{c}{$72^\circ<\theta_{1,2}<90^\circ$}  \\ 
Setup & Modules   & $\left<\Delta\phi\right>$  & $\mu$ & $\left<\Delta\phi\right>$ & $\mu$  & $\left<\Delta\phi\right>$ & $\mu$   \\
\hline
1 & GaGG-1.9 & 15.3$^\circ$ & $0.34\pm0.02$ &  15.3$^\circ$ & $0.34\pm0.01$&  15.3$^\circ$ & $0.30\pm0.01$ \\
2 & GaGG-2.9 & 18.3$^\circ$ & $0.29\pm0.02$ & 18.2$^\circ$ & $0.29\pm0.01$&  18.2$^\circ$ & $0.26\pm0.01$ \\
3 & GaGG-3.0 & 18.9$^\circ$ & $0.30\pm0.03$ &  18.9$^\circ$ & $0.30\pm0.01$&  19.0$^\circ$ & $0.29\pm0.01$\\
4 & LYSO-2.0 & 16.9$^\circ$ & $0.34\pm0.02$ &  16.9$^\circ$ & $0.33\pm0.01$&  16.7$^\circ$ & $0.31\pm0.01$\\
5 & LYSO-1.9 & 15.8$^\circ$ & $0.33\pm0.01$ &  15.8$^\circ$ & $0.32\pm0.01$&  15.8$^\circ$ & $0.31\pm0.01$\\ \hline

\end{tabular}
\end{table}

\section{Results and discussion}
We compare modulation factors obtained with the five detector setups. Three nominal angular ranges ($\pm2^\circ,\pm5^\circ, \pm10^\circ$) are selected around the scattering angle $\theta_{1,2}=82^\circ$, where the maximum correlation of azimuthal angles $\phi_1$ and $\phi_2$ is expected. In addition, we compare the modulation dependence on the angular resolution, $\Delta \theta$ and the mean azimuthal resolution $\left<\Delta\phi\right>$. 

\begin{figure}
	\centering
		\includegraphics[width=10cm]{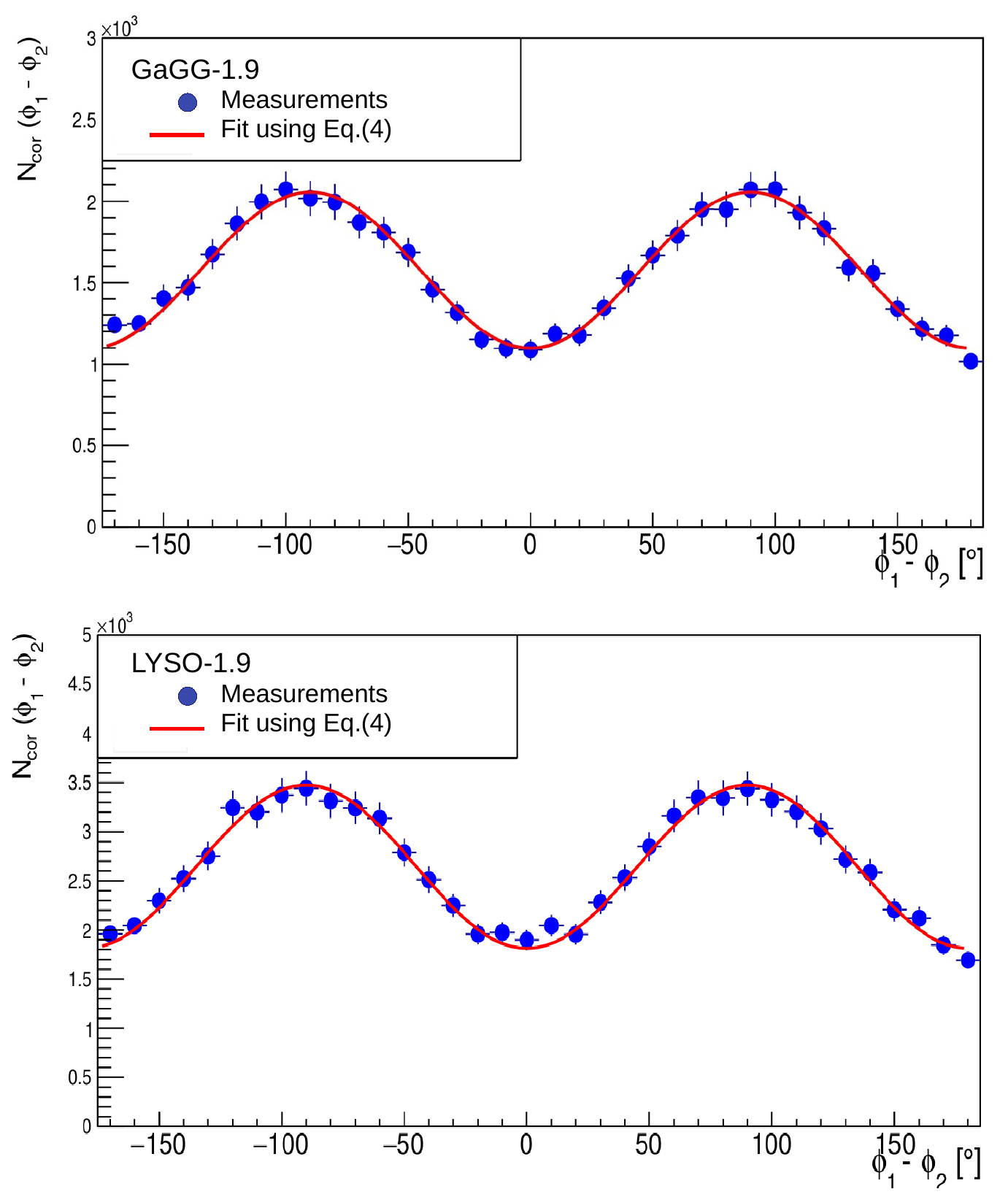}
	\caption{Observed azimuthal angle difference distributions for GaGG-1.9 and LYSO-1.9 detector configurations for $72^\circ<\theta_{1,2}<90^\circ$ angles and inter-pixel distance $d_{1,2}\!>\!4.5$ mm.}
	\label{FIG:3}
\end{figure}

Examples of the determined $\mathrm{N_{cor}(\phi_1-\phi_2)}$ distributions for GaGG-1.9 and LYSO-1.9 configurations are shown in Figure \ref{FIG:3}. The results show the modulation of the $\mathrm{N(\phi_1-\phi_2)}$ distribution, with maxima at $\pm 90^\circ$, as expected due to initial orthogonality of polarizations of the annihilation quanta. The subset of the results obtained by selecting events with $d_{1,2}\!>\!4.5$ mm for all configurations are shown in Table \ref{tbl2}. These results generally show that the strength of the modulation increases for scattering ranges closer to the $\theta_{1,2}=82^\circ$. This is expected, since the smaller $\theta$ window will allow the cross section (Eqn. 1.) to contribute in a narrow range closer to its maximum. In addition, it is clear that   higher modulations are reachable with more finely segmented crystals (GaGG-1.9, LYSO-1.9, LYSO-2.0) which can achieve lower $\left<\Delta\phi\right>$ compared to the larger crystal configurations (GaGG-2.9, GaGG-3.0).

\begin{figure}
	\centering
		\includegraphics[width=12cm]{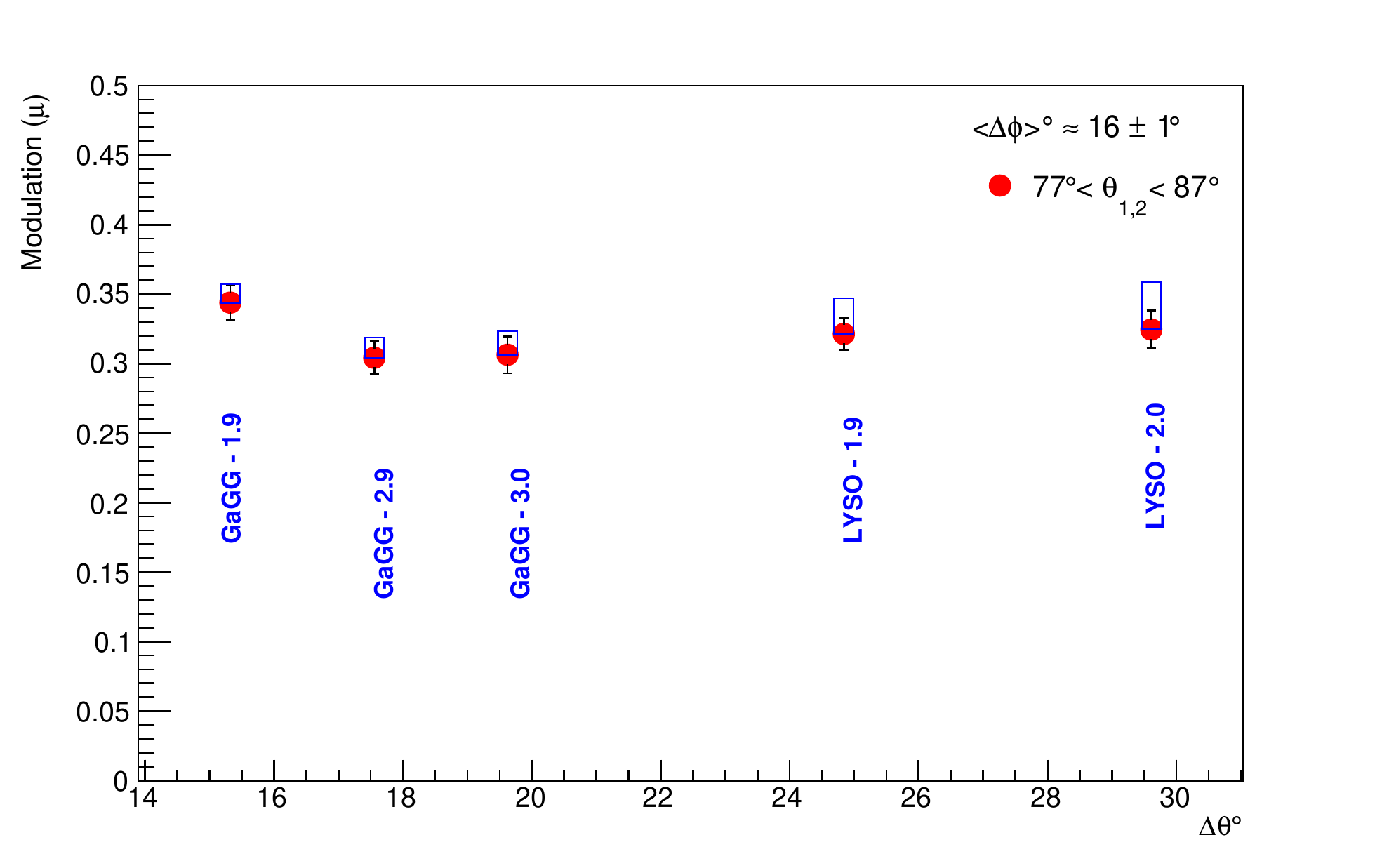}
	\caption{Modulation amplitude vs. $\Delta\theta$ at scattering angles $77^\circ\!<\!\theta_{1,2}\!<\!87^\circ$ and $\left<\Delta\phi\right>\!\approx\!16\pm1^\circ$ for all configurations. Systematic uncertainties are shown as blue frames.}
	\label{FIG:4}
\end{figure}

The values of $\mu$ listed in Table \ref{tbl2} are additionally subject to systematic uncertainties since the effective width of the $\theta$ window is larger than nominal due to the uncertainty $\sigma_\theta$. Since the maximum modulation is expected at $\theta_{1,2}=82^\circ$, the systematic uncertainties quantify by how much the measured value under-estimates the average $\mu$ in the nominal theta window. For $\mathrm{80^\circ\!<\!\theta_{1,2}<84^\circ}$, they range from 2.7\% to 7.7\%, for the setups 1-5, respectively. For $\mathrm{77^\circ\!<\!\theta_{1,2}<87^\circ}$, they range from 3.9\% to 10.5\% and for $\mathrm{72^\circ\!<\!\theta_{1,2}<90^\circ}$, they range from 5.8\% to 14.3\%, for the setups 1-5, respectively. The example in Figure \ref{FIG:4} shows that the energy resolution and consequently $\Delta \theta$ resolution do not critically influence  the amplitude of the $\mu$, at least for the applied selection criteria, they rather limit the precision by which it can be determined. The three configurations with comparable segmentation (GaGG-1.9, LYSO-1.9, LYSO-2.0) demonstrate similar polarimetric performance, in-spite of the significant differences between the energy resolution of GaGG and LYSO crystals.   

\begin{figure}
	\centering
		\includegraphics[width=12cm]{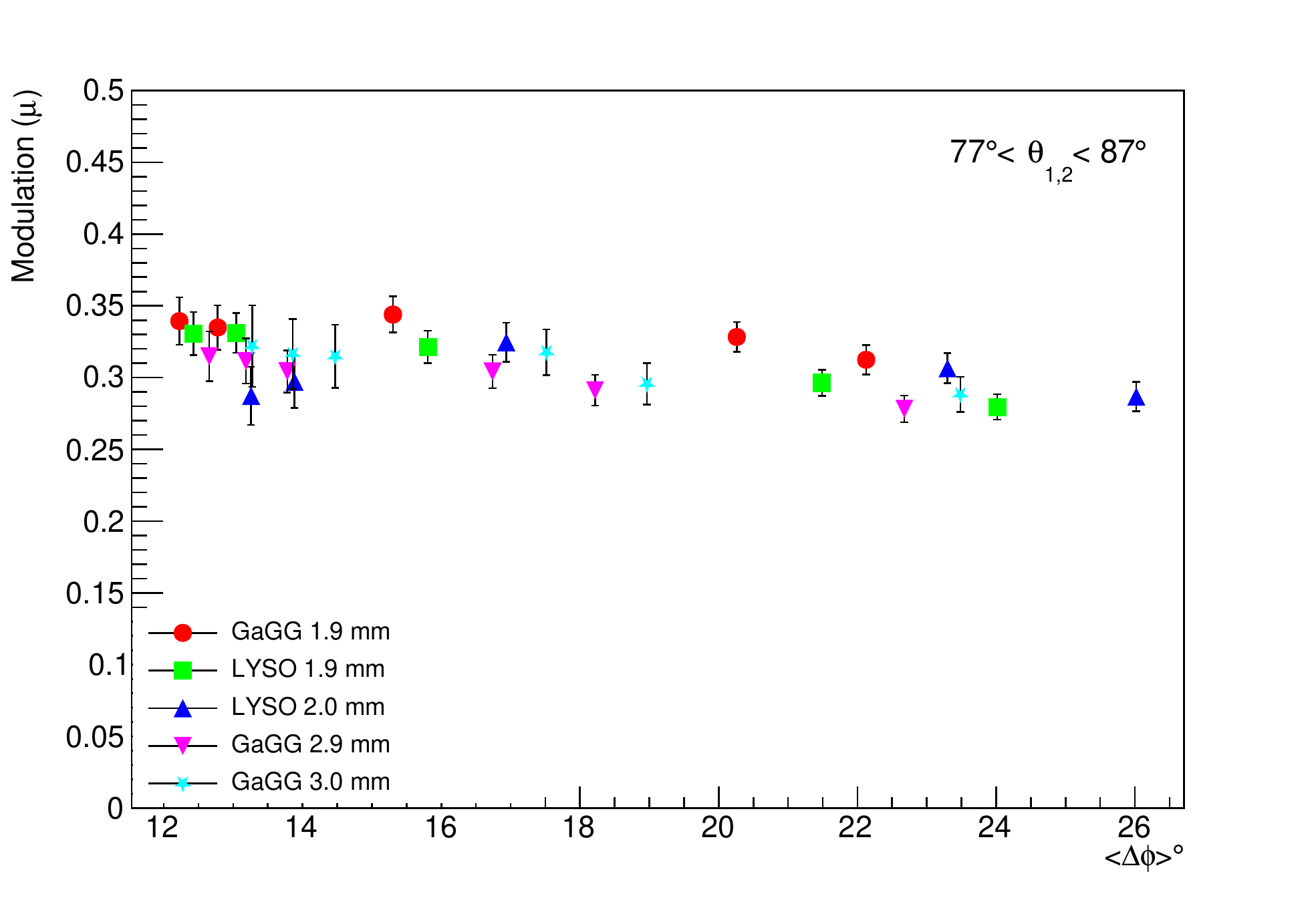}
	\caption{Modulation amplitude vs. mean azimuthal resolution $\left<\Delta\phi\right>$ at scattering angles $77^\circ\!<\!\theta_{1,2}\!<\!87^\circ$ for all configurations. Only statistical errors are shown. }
	\label{FIG:5}
\end{figure}

For a better understanding of the dependence of modulation factors on the azimuthal resolution, $\mu$ is plotted against $\left<\Delta\phi\right>$ in Figure \ref{FIG:5} for $d_{min}$ ranging from 3 mm to 36 mm, resulting in the mean values of $\left<\Delta\phi\right>$ from $12^\circ$ to $27^\circ$ (FWHM). An increasing trend in the modulation amplitude with better azimuthal resolutions is observed. The results strongly suggest that the dominant factor driving the polarimetric performance is the azimuthal resolution, $\left<\Delta\phi\right>$ resulting from the detector segmentation rather than the scattering angle resolution $\Delta\theta$ resulting from the energy resolution of the crystal. 

\section{Conclusions}
Angular correlations of annihilation quanta are successfully measured with single-layer segmented scintillation detectors. Polarimetric performance of five different detector configuration has been tested with crystals of either GaGG:Ce or LYSO:Ce and pixels sides ranging from 1.9 – 3.0 mm. All configurations can successfully reconstruct Compton events and measure the azimuthal modulation. The largest modulation amplitude is observed with more finely segmented crystals, with both materials, GaGG and LYSO performing comparatively in the selected regime. It is confirmed that the modulation amplitude depends on the mean azimuthal resolution resulting from detector segmentation, while no definite dependence on the energy resolutions of the selected crystals is observed. The present work confirms the feasibility of using the  single-layer detectors to measure the azimuthal correlations of annihilation quanta. This concept may be utilized to improve the sensitivity of next-generation clinical PET systems without a significant increase in hardware complexity or cost. 

\section{Acknowledgements}
This work was supported by the “Research Cooperability” Program of the Croatian Science Foundation, funded by the European Union from the European Social Fund under the Operational Programme Efficient Human Resources 2014–2020, grant number PZS-2019-02-5829.


\newpage

\bibliographystyle{unsrt}

\end{document}